\documentclass{article}

\usepackage{amsmath}
\usepackage{amsfonts}
\usepackage{stmaryrd}
\usepackage{amssymb}
\usepackage{amsthm}
\usepackage{color}
\usepackage{graphicx}
\usepackage{verbatim}
\usepackage{url}
\usepackage{hyperref}
\usepackage{authblk}
\usepackage{cite}

\DeclareMathOperator{\Sym}{Sym}

\newcommand{\vx}{\mathbf{x}}
\newcommand{\vy}{\mathbf{y}}
\newcommand{\vz}{\mathbf{z}}

\newcommand{\vone}{\mathbf{1}}

\newcommand{\cM}{\mathcal{M}}
\newcommand{\vzero}{\mathbf{0}}

\renewcommand{\epsilon}{\varepsilon}

\newtheorem{theorem}{Theorem}

\newtheorem{lemma}{Lemma}

\theoremstyle{definition}
\newtheorem*{definition}{Definition}

\newtheorem*{fix}{Fixation Axiom}

\numberwithin{equation}{section}

\begin{document}

\title{Symmetry in models of natural selection}
\author{Benjamin Allen}
\renewcommand\Affilfont{\itshape\small}
\affil{Department of Mathematics, Emmanuel College, Boston, MA, USA\\ORCID:0000-0002-9746-7613}
\date{}

\maketitle

\begin{abstract}
    Symmetry arguments are frequently used---often implicitly---in mathematical modeling of natural selection. Symmetry simplifies the analysis of models and reduces the number of distinct population states to be considered. Here, I introduce a formal definition of symmetry in mathematical models of natural selection. This definition applies to a broad class of models that satisfy a minimal set of assumptions, using a framework developed in previous works. In this framework, population structure is represented by a set of sites at which alleles can live, and transitions occur via replacement of some alleles by copies of others. A symmetry is defined as a permutation of  sites that preserves probabilities of replacement and mutation. The symmetries of a given selection process form a group, which acts on population states in a way that preserves the Markov chain representing selection. Applying classical results on group actions, I formally characterize the use of symmetry to reduce the states of this Markov chain, and obtain bounds on the number of states in the reduced chain.
\end{abstract}

\section{Introduction}

\begin{figure}
    \centering
    \includegraphics[width=\textwidth]{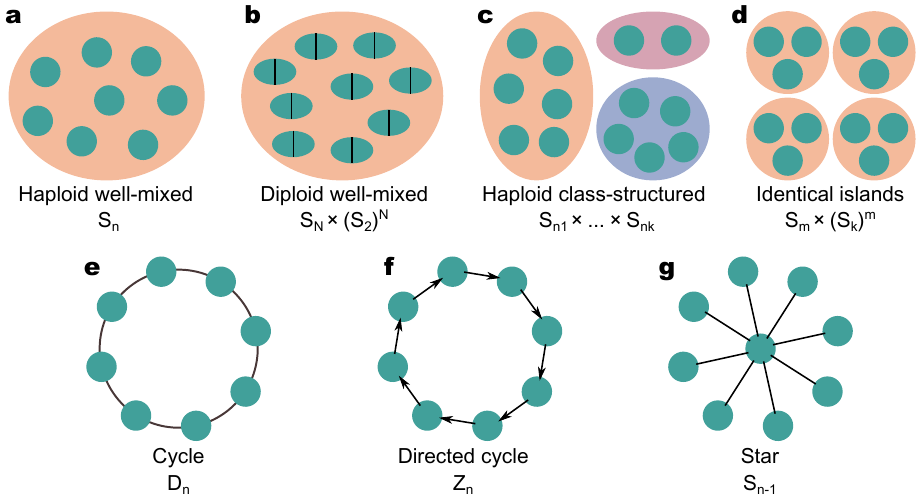}
    \caption{\textbf{Population structures and their symmetry groups. a} In a haploid well-mixed population model, such as the Moran \cite{Moran} and Wright-Fisher \cite{fisher1930genetical,wright1931evolution} models,  every permutation of the set $G$ of sites is a symmetry.  The symmetry group $\Sym(G,p)$ is therefore isomorphic to the symmetric group $S_n$, where $n=|G|$ is the number of sites. \textbf{b} In a diploid well-mixed population of size $N$, symmetries may permute individuals, and may also swap the two sites within individuals; the symmetry group is therefore (isomorphic to) $S_N \times (S_2)^N$. \textbf{c} In a haploid class-structured population, where the classes have respective sizes $n_1, \ldots, n_k$, symmetries may permute the members of each class. The symmetry group is therefore $S_{n_1} \times \ldots \times S_{n_k}$. \textbf{d} For a population structured as a cycle graph, symmetries may rotate and/or reflect the cycle; the symmetry group is the dihedral group $D_n$. \textbf{e} For a \emph{directed} cycle, only rotations are allowed;  the symmetry group is the cyclic group $Z_n$. \textbf{e} For a star graph, any permutation of the $n-1$ leaves is a symmetry, yielding the symmetry group  $S_{n-1}$.}
    \label{fig:examples}
\end{figure}

Natural selection is a complex process.  Even in the relatively simple case of two alleles competing in a fixed environment, heterogeneity among individuals can lead to a large number of possible population states, presenting a challenge for mathematical modeling. To make analytical progress, mathematical models typically assume some form of symmetry (Fig.~\ref{fig:examples}), which reduces the number of distinct states that need be considered.  

The simplest models of natural selection \cite{haldane1924mathematical,fisher1930genetical,wright1931evolution,Moran} assume that  individuals with the same genotype are interchangeable. Under this strong form of symmetry, one needs only keep track of the number of individuals with each genotype.

These models can refined to take into account individual differences based on sex \cite{wright1921systems,fisher1930genetical}, age \cite{fisher1930genetical,CharlesworthAge}, and other factors, collectively referred to as ``class"  \cite{taylor1990allele,lion2018class,lehmann2020individuals}. In class-structured models of selection, one typically assumes that individuals of the same class and genotype are interchangeable. Consequently, one needs only track the abundance of each genotype within each class.

Other models incorporate spatial structure. The earliest models of selection in spatially structured populations, such as Wright's island model \cite{WrightIsolation} and lattice models \cite{kimura1964stepping,NowakMay,LionvanBaalen}, also possessed a high degree of symmetry. More recently, attention has turned to asymmetric spatial structures. These are usually represented as graphs, in which each vertex corresponds to an individual and edges represent spatial or social relationships \cite{ErezGraphs,SantosScaleFree,Ohtsuki,debarre2014social,allen2017evolutionary,moller2019exploring,su2022asymmetric,yagoobi2023categorizing}.

A symmetry can be mathematically represented as a permutation---a bijective mapping from a set to itself---that preserves some relevant structure.  For example, a symmetry of a graph (also called a graph automorphism) is a permutation $\sigma$ of the vertex set that preserves edges---meaning that vertices $i$ and $j$ are joined by an edge if and only if $\sigma(i)$ and $\sigma(j)$ are as well. The symmetries of a given object form a group under composition of permutations (see, e.g., \cite{armstrong1997groups}). 

Applying this mathematical notion of symmetry, Taylor \cite{Taylor}  studied selection on graphs possessing a form of symmetry called bi-transitivity. This was followed by other investigations of selection on graphs with various symmetry properties  \cite{grafen2007inclusive,taylor2011groups,AllenGraphMut,debarre2014social,taylor2014hamilton}. An overarching  theory of symmetry for selection on graphs was developed by
McAvoy and Hauert \cite{mcavoy2015structural}, who showed how graph symmetries preserve properties of the Markov chain representing selection. 

In an effort to unify the wide variety of models of natural selection in structured populations, my collaborators and I have proposed a general mathematical modeling framework for natural selection \cite{allen2014measures,allen2019mathematical}. This framework defines a class of models of natural selection, which includes the classical Moran \cite{Moran} and Wright-Fisher \cite{fisher1930genetical,wright1931evolution} models, game-theoretic processes in finite populations \cite{NowakFinite,imhof2006evolutionary,LessardFixation},  selection processes on graphs \cite{ErezGraphs,SantosScaleFree,Ohtsuki,debarre2014social,allen2017evolutionary,moller2019exploring,su2022asymmetric,yagoobi2023categorizing}, and models of haplodiploid social insect colonies \cite{allen2019mathematical}. For  models in this framework, selection occurs between two alleles at a single genetic locus, in a population of fixed size. The population structure is represented by a fixed set of sites at which alleles can live. A given population state is identified by specifying which allele occupies each state. Natural selection is represented as a Markov chain, in which each transition involves the replacement of some alleles by others, possibly with mutation. Using this framework, general results---applicable to any model in the class---have been derived regarding fixation probabilities \cite{allen2015molecular,mcavoy2021fixation,belanger2023asymmetric}, allele frequencies \cite{mcavoy2018stationary}, social behavior \cite{McAvoy2020EvaluatingTS}, and genetic assortment \cite{Allen2022Coalescent}.

Here, using this framework, I develop a mathematical theory of symmetry in models of natural selection. A symmetry will be defined as a permutation of sites that preserves probabilities of replacement and mutation.  The symmetries of a given natural selection process form a group under composition. This group acts on the set of sites, and by extension, on the set of possible population states. 

Applying this theory, I formalize the use of symmetry to reduce the size of the Markov chain representing selection. Two states are equivalent if there is a symmetry transforming one into the other. By grouping equivalent states together, one obtains a reduced Markov chain, the states of which are orbits of the group action. Using classical results on orbit counting, I derive upper and lower bounds for the number of states in the reduced chain.

Section \ref{sec:framework} summarizes the relevant modeling framework. Section \ref{sec:symmetry} introduces the formal definition of symmetry.  This definition is used in Section \ref{sec:equivsites} to formalize the notion of classes of sites, and in Section \ref{sec:reduced} to show how symmetry reduces the Markov chain representation of selection.

\section{Modeling framework}
\label{sec:framework}

\begin{figure}
\centering
\includegraphics{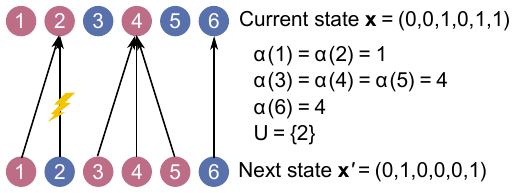}
\caption{\textbf{Modeling framework.} 
The modeling framework used here \cite{allen2014measures,allen2019mathematical} defines a class of models with two competing alleles (numbered 0 and 1) at a single genetic locus. There a fixed set $G$ of sites at which alleles can live; here, $G=\{1,2,3,4,5,6\}$. Haploids (pictured here) have one site per individual; diploids two. The population state is a binary vector $\vx=(x_g)_{g \in G}$ specifying which allele occupies each site.  Each time-step, a parentage map $\alpha$ and mutation set $U$ are sampled from a joint probability distribution that depends on the current state $\vx$. To form the next state, $\vx'$, each non-mutated site $g \notin U$ inherits the allele of the parent, $x'_g = x_{\alpha(g)}$, while each mutated site $g \in U$ inherits the opposite allele, $x'_g = 1-x_{\alpha(g)}$. This leads to a Markov chain $\cM(G,p)$ representing  natural selection.}
\label{fig:Framework}
\end{figure}

I begin by reviewing the mathematical modeling framework---developed in Refs.~\cite{allen2014measures,allen2019mathematical} and depicted in Figure \ref{fig:Framework}---to which the proposed definition of symmetry will apply. This framework provides a common notation and foundational assumptions for a class of models, each representing selection between two alleles at a single genetic locus.

\subsection{Sites, alleles, and states}

There are two allele types, numbered 0 and 1. Taking a gene's-eye view, the population structure is represented by a finite set $G$ of $n$ genetic sites, at which alleles live.  Each site houses one allele, and each individual contains a number of alleles equal to its ploidy (one for haploids, two for diploids). 

The allele occupying site $g \in G$ is represented by the variable $x_g \in \{0,1\}$.  These variables are combined into a $g$-indexed binary vector $\vx=(x_g)_{g \in G} \in \{0,1\}^G$ representing the population state.

\subsection{Replacement and mutation}

State transitions occur via replacement (of some alleles by copies of others) and  mutation. Replacement is represented by a set mapping $\alpha:G \to G$, called a \emph{parentage map}. For each site $g\in G$, $\alpha(g)$ indicates the site from which the new allele in $g$ is either survived or inherited. In other words, $\alpha(g)=h$ can mean either that the allele in $g$ died and was replaced by a copy of that in $h$, or else that the allele in $h$ survived and moved to $g$ (or stayed in $g$, if $g=h$). These two possibilities are formally equivalent, in that both result in site $h$ transmitting its allele to $g$.

Mutation occurs after replacement, and is represented by a subset $U \subseteq G$ of sites that acquire mutations.  Mutation interchanges alleles 0 and 1.

In each state $\vx$, the joint probability that  parentage map $\alpha$ and mutation set $U$ occur is denoted $p_\vx(\alpha,U)$.  This probability $p$ may be understood as a real-valued function of three arguments,  $p_\cdot(\cdot,\cdot)$, such that, for each $\vx \in \{0,1\}^G$, $p_\vx(\cdot,\cdot)$ is a joint probability distribution over all possible set mappings $\alpha:G\to G$ and subsets $U \subseteq G$.

\subsection{Fixation axiom}

For the population to be unitary, at least one site should be able (with positive probability) to eventually spread its descendants to all sites. This principle is formalized as follows \cite{allen2019mathematical}:
\begin{fix}
    There exists a site $g \in G$, and a finite sequence of parentage maps $\alpha_1, \ldots,\alpha_m$, such that 
    \renewcommand{\labelenumi}{(\roman{enumi})}
    \begin{enumerate}
        \item $p_\vx(\alpha_k)>0$ for all $1\le k \le m$ and $\vx \in \{0,1\}^G$, and 
        \item For all $h \in G$, $\alpha_1\circ \cdots \circ \alpha_m(h)=g$.
    \end{enumerate}
\end{fix}

A set of sites $G$ and probability function $p=p_\cdot(\cdot,\cdot)$, satisfying the Fixation Axiom, define a \emph{selection process}. A selection process $(G,p)$ captures all structures and processes relevant to selection, including behavioral interaction, spatial structure, migration, and mating pattern; these are all represented implicitly in the probability function $p$.

\subsection{Selection Markov chain}

Given a selection process $(G,p)$, one can construct a Markov chain $\cM(G,p)$, which I call the \emph{selection Markov chain}, on the set of population states $\{0,1\}^G$. From a given state $\vx$, the subsequent state $\vx'$ is determined by first sampling a parentage map $\alpha$ and mutation set $U$ from the probability distribution $p_\vx(\cdot,\cdot)$, and then setting
\begin{equation}
x'_g = \begin{cases}
    x_{\alpha(g)} & \text{if $g \notin U$}\\
    1-x_{\alpha(g)} & \text{if $g \in U$.}
\end{cases}
\end{equation}
In this way, each non-mutated site $g \notin U$ inherits the allele from its parent site $\alpha(g)$, while each mutated site $g \in U$ inherits the \emph{opposite} allele from its parent site.

\subsection{Scope and limitations}

This framework encompasses a wide variety of models of natural selection, including classical models of well-mixed populations \cite{fisher1930genetical,wright1931evolution,Moran,cannings1974latent}, models of frequency-dependent selection \cite{NowakFinite,imhof2006evolutionary,LessardFixation}, and models with class \cite{taylor1990allele}, island \cite{wright1931evolution,WrightIsolation}, spatial \cite{nowak1993spatial,LionvanBaalen}, network \cite{ErezGraphs,SantosScaleFree,Ohtsuki,debarre2014social,allen2017evolutionary,moller2019exploring,su2022asymmetric} and/or mating \cite{wright1921systems,belanger2023asymmetric} structure.  However, this framework does not include models where the population size or structure can vary, such as models of selection on dynamic graphs \cite{PachecoCoevolution,wu2010evolution,Su2023StrategyEO}.

\section{Symmetries of a selection process}
\label{sec:symmetry}

We are now prepared to define the symmetries of a model encompassed by this framework.

\subsection{Definition}

The definition of symmetry employs the following notation: For any state $\vx \in \{0,1\}^G$ and set mapping $\tau:G \to G$, let $\vx_\tau \in \{0,1\}^G$ denote the state that has allele $x_{\tau(g)}$ in each site $g \in G$; that is, $(\vx_\tau)_g=x_{\tau(g)}$. To illustrate the usefulness of this notation, observe that if parentage map $\alpha$ occurs in state $\vx$, and there is no mutation ($U=\varnothing$), the subsequent state is then $\vx'=\vx_\alpha$.  This notation obeys a composition rule: if $\sigma,\tau:G \to G$ are two set mappings, then $(\vx_\sigma)_\tau=\vx_{\sigma \circ \tau}$.

With this notation, symmetry is defined as follows:

\begin{definition} A \emph{symmetry} of a selection process $(G,p)$ is a permutation $\sigma:G \to G$ such that, for every state $\vx$, parentage map $\alpha:G \to G$, and mutation set $U \subseteq G$,
\begin{equation}
\label{eq:symmetry}
p_\vx(\alpha,U) = p_{\vx_\sigma} \big(\sigma^{-1} \circ \alpha \circ \sigma, \sigma^{-1}(U) \big).
\end{equation}
\end{definition}

\begin{figure}
    \centering
    \includegraphics[width=\textwidth]{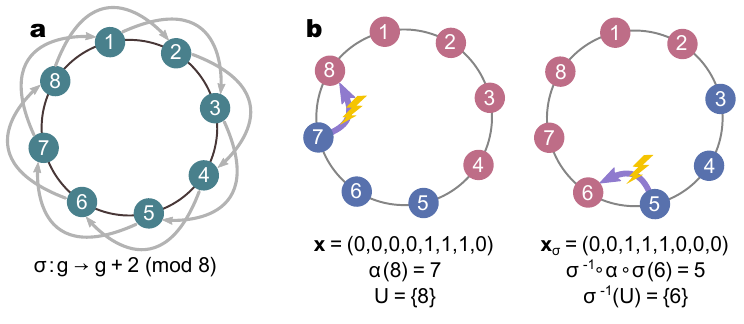}
    \caption{\textbf{Symmetry on a cycle.  a.} A population is structured as a cycle graph of size 8. For ease of reference, the sites here are numbered $g=1,\ldots,8$.  (In general, sites need not be numbered.) The group of symmetries is the dihedral group $D_8$. The symmetry  $\sigma:g \to g+2\pmod{8}$ is illustrated here.  \textbf{b.} An illustration of the symmetry condition, Eq.~\eqref{eq:symmetry}. In the state $\vx$ depicted, site 8 is replaced by mutated offspring of site 7. This is represented by a parentage map with $\alpha(8)=7$ (and $\alpha(g)=g$ for all $g \ne 8$) and $U=\{8\}$.  As shown at right, this is equivalent to parentage map $\sigma \circ \alpha \circ \sigma^{-1}$ and mutation set $\sigma^{-1}(U)$ occurring in state $\vx_\sigma$.}
    \label{fig:cycle_example}
\end{figure}

Figure \ref{fig:cycle_example} provides an example to illustrate Eq.~\eqref{eq:symmetry}. The idea is that, if $\sigma$ is a symmetry, then any state $\vx$ is equivalent to its permuted state $\vx_\sigma$, and every transition from state $\vx$ has a corresponding equivalent in state $\vx_\sigma$. If, in a transition from state $\vx$, site $g$ has parent $\alpha(g)=h$, the equivalent transition in state $\vx_\sigma$ has $\alpha(\sigma(g))=\sigma(h)$, or equivalently, $\sigma^{-1}\circ \alpha \circ \sigma(g)=h$. Likewise, if in a transition from state $\vx$, site $g$ acquires a mutation ($g \in U$), the equivalent transition in state $\vx_\sigma$ has $\sigma(g) \in U$, or $g \in \sigma^{-1}(U)$.  So overall, the pair $(\alpha,U)$ occurring in state $\vx$ is equivalent to the pair  $(\sigma^{-1}\circ \alpha \circ \sigma, \sigma^{-1}(U))$ occurring in state $\vx_\sigma$, which motivates Eq.~\eqref{eq:symmetry}. These arguments are made precise in the proof of Theorem \ref{thm:sym_Markov} in Appendix \ref{app}. 

\subsection{Symmetry group and its actions}

The symmetries of a given selection process $(G,p)$ form a group under composition of permutations.  This is proven as Lemma \ref{lem:group} in Appendix \ref{app}. We denote this group $\Sym(G,p)$. Figure \ref{fig:examples} shows the symmetry groups for various models of structured populations.

Since each symmetry is a permutation, $\Sym(G,p)$ is a subgroup of the symmetric group $\Sym(G)$---that is, the group of all permutations of $G$. For a haploid, well-mixed population, all permutations are symmetries, so $\Sym(G,p) =\Sym(G)$.  For models with additional structure (space, sex, class, graph, etc.), $\Sym(G,p)$ is the subgroup of $\Sym(G)$ that preserves this structure, i.e.~the subgroup of permutations satisfying Eq.~\eqref{eq:symmetry}. 

Two group actions of $\Sym(G,p)$ are relevant to the role of symmetry in selection processes.  The first is the left group action of $\Sym(G,p)$ on the set of sites $G$, given by $g \mapsto \sigma(g)$. The second is the right group action of $\Sym(G,p)$ on the set of states $\{0,1\}^G$ given by $\vx \mapsto \vx_\sigma$. As will be shown later, these group actions help to formalize the symmetry arguments that often arise in analyses of models of natural selection. 

\subsection{Examples}

Without going into full modeling details, it is helpful to identify the kinds of symmetry that arise in different varieties of evolutionary models:

\begin{itemize}
    \item \emph{Well-mixed haploid populations} (Fig.~\ref{fig:examples}a): The simplest models of selection, such as the Moran \cite{Moran}, haploid Wright-Fisher \cite{fisher1930genetical,wright1931evolution}, and Cannings exchangeable \cite{cannings1974latent} models---as well as frequency-dependent generalizations thereof \cite{NowakFinite,imhof2006evolutionary,LessardFixation}---describe a well-mixed, haploid population.  In this case, each genetic site corresponds to an individual, and any two genetic sites are interchangeable.  For such models, any permutation of the set $G$ of sites is a symmetry. The symmetry group is $\Sym(G)$, which is isomorphic to $S_n$, the symmetry group on $n=|G|$ elements.
    \item \emph{Well-mixed diploid populations} (Fig.~\ref{fig:examples}b): A well-mixed hermaphroditic diploid population can be represented by specifying a finite set $I$ of individuals and a partition of the set of sites, $G = \bigsqcup_{i \in I} G_i$, where each subset $G_i$ contains two sites to house the two alleles in individual $i$. The diploid Wright-Fisher model \cite{fisher1930genetical,wright1931evolution}, for example, may be represented this way \cite{Allen2022Coalescent}. For such models, any permutation $\sigma$ that preserves the partition into individuals---meaning that $\{ \sigma(G_i)\}_{i\in I} = \{ G_i\}_{i\in I}$---is a symmetry.  A symmetry may interchange individuals, and also may interchange the two alleles within an individual. This is because an individual's two alleles are generally considered interchangeable ($Aa$ and $aA$ genotypes are equivalent), unless the alleles are sex-linked. Letting $N=|I|=n/2$ denote the number of individuals, $\Sym(G,p)$ is isomorphic to $S_N \times (S_2)^N$.
    \item \emph{Class-structured populations} (Fig.~\ref{fig:examples}c): In a class-structured population \cite{taylor1990allele}, individuals are distinguished by sex or by other factors such as role within a colony. Class structure in a haploid population can be represented by a partition $G = \bigsqcup_{j=1}^k C_j$, with each subset $C_j$ representing the sites in individuals of class $j$. Models of class-structured populations typically assume that individuals with the same class and genotype are indistinguishable. In this case, a symmetry is a permutation $\sigma$ that leaves each class fixed, in the sense that $\sigma (C_j) =C_j$ for each $j=1,\ldots, k$. $\Sym(G,p)$ is therefore isomorphic to $S_{n_1} \times \cdots \times S_{n_k}$, where $n_j=|C_j|$ is the number of indivdiuals in class $j$.
    \item \emph{Island-structured populations} (Fig.~\ref{fig:examples}d): In island models \cite{wright1931evolution,WrightIsolation,taylor1992altruism,lehmann2016invasion} and metapopulation models \cite{levins1969some,hanski1998metapopulation,wade2019adaptation}, the population is divided into subpopulations (``islands"), which are joined to each other by low levels of migration. Within each island, individuals with the same genotype are considered indistinguishable. In our framework, assuming haploid genetics, this kind of model is again represented using a partition $G= \bigsqcup_{j=1}^m G_j$, where here $j=1,\ldots,m$ index the islands.  It is often assumed that the islands have equal size and are interchangeable with each other. In this case, a symmetry is any permutation that preserves the partition into islands; that is $\{\sigma(G_j)\}_{j=1}^m=\{G_j\}_{j=1}^m$. Note that symmetries may interchange islands with each other, and may also interchange the sites within any island. $\Sym(G,p)$ is therefore isomorphic to $S_m \times (S_k)^m$, where $k$ is the size of each island, i.e.~$k=|G_j|$ for each $j=1,\ldots,m$.
    \item \emph{Graph-structured populations} (Fig.~\ref{fig:examples}e--g): There are many models of natural selection on graphs \cite{ErezGraphs,SantosScaleFree,Ohtsuki,debarre2014social,allen2017evolutionary,moller2019exploring,su2022asymmetric,yagoobi2023categorizing}, including weighted graphs \cite{debarre2014social,allen2017evolutionary}, directed graphs \cite{su2022asymmetric}, and multilayer networks \cite{su2022multilayer}.  Although these models differ in their details, they share the property that any symmetry (automorphism) of the graph is also a symmetry of the selection process. In this case, $\Sym(G,p)$ is the automorphism group of that graph; see McAvoy and Hauert \cite{mcavoy2015structural} for closely related results.
\end{itemize}

\subsection{Symmetry and transition probabilities}

If the proposed definition of symmetry is to be useful, symmetries should preserve the transition probabilities of the selection Markov chain $\cM(G,p)$. Denoting the transition probability from state $\vx$ to state $\vy$ by $P_{\vx \to \vy}$, this principle is formalized in the following theorem (proven in Appendix \ref{app}):

\begin{theorem}
\label{thm:sym_Markov}
For any symmetry $\sigma \in \Sym(G,p)$ of a selection process $(G,p)$, and any states $\vx,\vy \in \{0,1\}^G$, $P_{\vx \to \vy} = P_{\vx_\sigma \to \vy_\sigma}$. 
\end{theorem}

Theorem \ref{thm:sym_Markov} guarantees that the action of $\Sym(G,p)$ on $\{0,1\}^G$ preserves the transition probabilities of $\cM(G,p)$ and, by exetension, all quantities derived from them. For example,  $m$-step transition probabilities are also preserved by symmetries, in the sense that $P^{(m)}_{\vx \to \vy} = P^{(m)}_{\vx_\sigma \to \vy_\sigma}$.  Likewise, if $\cM(G,p)$ admits a unique stationary distribution $\pi$ (which occurs whenever there is complete gene flow and recurring mutation; see Theorem 1 of Ref.~\cite{allen2014measures}), then $\pi(\vx)=\pi(\vx_\sigma)$ for any state $\vx$ and symmetry $\sigma$. This connects our definition of symmetry to the notion of ``evolutionary equivalence" introduced by McAvoy and Hauert \cite{mcavoy2015structural}.

\section{Equivalent sites}
\label{sec:equivsites}

Theoretical analyses of natural selection often invoke the idea that certain individuals or locations are equivalent within a given model. We can formalize this idea by defining an equivalence relation $\sim$ on on the sites $G$ of a given selection process $(G,p)$. Two sites $g,h \in G$ are equivalent,  $g \sim h$, if and only if there is a symmetry $\sigma \in \Sym(G,p)$ such that $\sigma(g)=h$. This equivalence relation induces a partition of $G$ into equivalence classes. We observe that these equivalence classes are the orbits of the group action of $\Sym(G,p)$ on $G$.  These equivalence classes formalize the notion of ``class" used in class-structured models of selection (Fig.~\ref{fig:examples}c).

If sites $g$ and $h$ are equivalent, then any property of site $g$ must hold for site $h$, and vice versa. 
As an example, consider fixation probability---the probability that a new mutant allele eventually spreads throughout a population \cite{patwa2008fixation,ErezGraphs,mcavoy2021fixation}.  For heterogeneously structured populations, the fixation probability depends on where the mutant allele originates \cite{maciejewski2014reproductive,adlam2015amplifiers,allen2015molecular,mcavoy2015structural}. However, if $g$ and $h$ are equivalent sites, mutations arising at these two sites should have the same fixation probability. To formalize this principle, we observe that if there is no mutation ($p_\vx(\alpha,U)=0$ unless $U=\varnothing$), the single-allele states $\vzero=(0, \ldots, 0)$ and $\vone=(1,\ldots,1)$ are absorbing in $\cM(G,p)$, and all other states are transient (see Theorem 2 of Ref.~\cite{allen2014measures}). Fix two equivalent sites $g,h\in G$ and a symmetry $\sigma \in \Sym(G,p)$ with $\sigma(g)=h$. Let $\vx$ be the state with allele 1 in site $h$ and 0's elsewhere; i.e., $x_h=1$ and $x_\ell=0$ for all $\ell \ne h$. Let $\rho_\vx$ denote the probability of eventual absorption in state $\vone$ from initial state $\vx$. Then, applying Theorem \ref{thm:sym_Markov}, we have
\[
\rho_\vx = \lim_{m \to \infty} P^{(m)}_{\vx \to \vone}
= \lim_{m \to \infty} P^{(m)}_{\vx_\sigma \to \vone_\sigma} 
= \lim_{m \to \infty} P^{(m)}_{\vx_\sigma \to \vone} = \rho_{\vx_\sigma}.
\]
This demonstrates that the fixation probability of allele 1 is the same whether starting from site $h$ (as it does in state $\vx$) or site $g$ (as it does in state $\vx_\sigma$). In short, symmetry preserves transition probabilities. This generalizes a result proved in Ref.~\cite{mcavoy2015structural} for models of selection on graphs.

An important special case arises when the action of $\Sym(G,p)$ on $G$ is \emph{transitive}, meaning that for every $g,h \in G$ there is some $\sigma \in \Sym(G,p)$ such that $\sigma(g)=h$.  For selection processes with transitive symmetry, all sites are equivalent. This formalizes the idea of a homogeneous population, in which any individual may be understood as representative of all. In Figure \ref{fig:examples}, the models depicted in panels a, b, d, e, and f possess transitive symmetry, while those in panels c and g do not. Transitive symmetry has been formally applied in models of selection on graphs \cite{Taylor,grafen2008natural,allen2014games,debarre2014social}, and is implicitly invoked in evolutionary arguments that reason from the point of view of an arbitrary ``focal individual" \cite{van2015social,Goodnight2013ONMS}.

\section{Reduction of the selection Markov chain}
\label{sec:reduced}

In models of natural selection, symmetry is used primarily as a way to reduce the number of states to be considered.  This is a useful principle, since a model with $|G|=n$ sites has $2^n$ possible states, which quickly becomes unwieldy.

\subsection{Construction of the reduced chain}

To formalize the reduction of the selection Markov chain by symmetry, observe that in light of Theorem \ref{thm:sym_Markov}, two states $\vx$ and $\vy$ can be considered equivalent if there is a symmetry $\sigma$ such that $\vy=\vx_\sigma$. This provides an equivalence relation $\simeq$ on the set of states $\{0,1\}^G$. 
The equivalence classes of $\simeq$ are the orbits of the group action of $\Sym(G,p)$ on $\{0,1\}^G$. We denote the equivalence class (orbit) of a state $\vx \in \{0,1\}^G$ by $[\vx]$:
\begin{equation}
    [\vx] = \{ \vx_\sigma \mid \sigma \in \Sym(G,p)\}.
\end{equation}
In particular, since the action of $\Sym(G,p)$ permutes alleles among sites, equivalent states must have the same number of each allele. Denoting the number of 1 alleles in state $\vx$ by $\Sigma \vx$, we have $\vx \simeq \vy \Rightarrow \Sigma \vx=\Sigma \vy$.

These orbits form the states of a reduced Markov chain, $\mathcal{R}(G,p)$.  The transition probability from $[\vx]$ to $[\vy]$ in $\mathcal{R}(G,p)$ is \begin{equation}
\label{eq:reduced}
    P_{[\vx] \to [\vy]} = \sum_{\vz \in [\vy]} P_{\vx \to \vz}.
\end{equation}
In light of Theorem \ref{thm:sym_Markov}, it does not matter which representative $\vx$ of the orbit $[\vx]$ is used in the right-hand side of Eq.~\eqref{eq:reduced}. This reduced chain $\mathcal{R}(G,p)$ simplifies the original chain $\cM(G,p)$ by grouping together all states that are equivalent by symmetry.

\subsection{Examples}

This idea of a reduced chain is  implicitly used in many established models:
\begin{itemize}
    \item For a well-mixed haploid population (Fig.~1a), two states $\vx$ and $\vy$ are equivalent if and only if they have the same number of 1 alleles: $\vx \simeq \vy \Leftrightarrow \Sigma \vx=\Sigma \vy$. The reduced chain $\mathcal{R}(G,p)$ therefore has $n+1$ states, which can be indexed $k=0,\ldots,n$ according to the number of 1 alleles. This corresponds to the standard representation of the Moran  and haploid Wright-Fisher  models as Markov chains on $\{0,\ldots,n\}$ (e.g.,~\cite{ewens2004mathematical}, Chapter 3).
    \item For a diploid hermaphroditic well-mixed population (Fig.~1b), we can regard the alleles $\{x_{i_1},x_{i_2}\}$ in each individual $i\in I$ as an unordered pair, representing its genotype. Two states are equivalent if and only if they have the same number of $\{0,0\}$, $\{0,1\}$, and $\{1,1\}$ genotypes. The states of the reduced chain $\mathcal{R}(G,p)$ can therefore be represented as triples $(k_{00},k_{01},k_{11})$ of nonnegative integers, subject to $k_{00}+k_{01}+k_{11}=n$.  There are $n(n+1)/2$ reduced states in total. Representations of this form have been used at least as far back as 1908 \cite{hardy1908mendelian}.
    \item For a class-structured population (Fig.~\ref{fig:examples}c) with classes of size $n_1, \ldots, n_k$, if any permutation that preserves classes is a symmetry, then two states are equivalent if they have the same number of 1 alleles in each class.  States of $\mathcal{R}(G,p)$ can be represented as $k$-tuples $(m_1, \ldots,m_k$), where $m_j$ is the number of 1 alleles in class $C_j$. There are $\prod_{j=1}^k(n_j+1)$ such reduced states. 
\end{itemize}

\subsection{Size of the reduced chain}

One may now ask, how much of a reduction is achieved? That is, how does the number of states in the reduced chain $\mathcal{R}(G,p)$ compare to the $2^{n}$ states of the original chain $\cM(G,p)$? 

\begin{figure}
    \centering
    \includegraphics{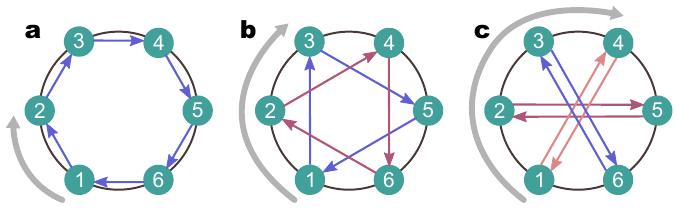}
    \caption{\textbf{The cycle decomposition of a rotation. a.} Rotating a cycle graph of size 6 by one vertex yields one cycle, which is denoted (1 2 3 4 5 6) in the notation of  cycle decompositions (e.g.,~\cite{armstrong1997groups}, Chapter 6). \textbf{b} Rotating by two vertices yields two cycles: (1 3 5) and (2 4 6). \textbf{c} Rotating by three vertices yields three cycles: (1 4), (2 5), and (3 6).  The cycle decomposition is used in Eq.~\eqref{eq:Polya} to count the states in the reduced Markov chain.  In general, for a directed cycle of size $n$ (Fig.~\ref{fig:examples}f), rotating by $j$ vertices creates $\gcd(n,j)$ cycles, yielding Eq.~\eqref{eq:reducedcycle}.}
    \label{fig:cycledecomp}
\end{figure}

This is equivalent to a well-known problem in combinatorics of finding the number of distinct colorings of a symmetric object \cite{polya1937kombinatorische,erickson2013introduction}. For us, the object is the set $G$ of sites and the colors correspond to alleles. P\'{o}lya's Enumeration Theorem \cite{redfield1927theory,polya1937kombinatorische,erickson2013introduction} provides a formula in terms of \emph{cycle decompositions} (Fig.~\ref{fig:cycledecomp}). Each symmetry $\sigma \in \Sym(G,p)$ partitions $G$ into nonoverlapping cycles, where sites $g,h \in G$ belong to the same cycle if and only if $g=\sigma^k(h)$ for some power $k$ of $\sigma$. (Here, $\sigma^k$ means the permutation $\sigma$ is  iteratively applied $k$ times.) Letting $c(\sigma)$ denote the number of cycles in the cycle decomposition associated to $\sigma$, and $S=|\Sym(G,p)|$ the number of symmetries, the number $R$ of reduced states is given by
\begin{equation}
\label{eq:Polya}
    R = \frac{1}{S}\sum_{\sigma \in \Sym(G,p)} 2^{c(\sigma)}.
\end{equation}
This formula generalizes to $m>2$ alleles by replacing 2 with $m$.

For the directed cycle (Fig.~\ref{fig:examples}f), $\Sym(G,p)$ consists of rotations by $j$ vertices, for $j=1,\ldots,n$. Such a rotation has $\gcd(n,j)$ cycles, where gcd denotes greatest common divisor (Fig.~\ref{fig:cycledecomp}). Applying Eq.~\eqref{eq:Polya}, the number of reduced states is
\begin{equation}
\label{eq:reducedcycle}
    R = \frac{1}{n}\sum_{j=1}^n 2^{\gcd(n,j)}.
\end{equation}
This equivalent to a known result for the ``necklace problem" in combinatorics \cite{riordan1957combinatorial,erickson2013introduction}, but expressed in a more elementary way.

In general, applying Eq.~\eqref{eq:Polya} requires computing the cycle decomposition of each symmetry, which is laborious if the number of symmetries is large. However, the following theorem (proven in Appendix \ref{app}) provides bounds on $R$  using only the numbers of sites and symmetries:

\begin{theorem}
For a selection process $(G,p)$, let $n=|G|$ be the number of states, $S=|\Sym(G,p)|$ be the number of symmetries, and $R$ be the number of states in the reduced chain $\mathcal{R}(G,p)$. Then $R$ is bounded by 
\begin{equation}
\label{eq:Rbounds}
    \max\left(n+1, 2 + \frac{2^{n}-2}{S} \right) \le R \le 2 + \frac{1}{S}\sum_{k=1}^{n-1} \binom{n}{k} \gcd(S,k!(n-k)!).
\end{equation}
\end{theorem}

The lower bound in Eq.~\eqref{eq:Rbounds} is really two lower bounds. First, there must be at least $n+1$ reduced states, one for each possible number $k=0,\ldots,n$ of alleles of type 1.  Second, the number of states cannot be reduced by a factor greater than the number $S$ of symmetries, and there is no reduction for the two single-allele states $\mathbf{0}$ and $\mathbf{1}$, which are invariant under permutation.  The upper bound in Eq.~\eqref{eq:Rbounds} comes from considering the minimal reduction that can be achieved in a state with $k$ alleles of type 1 and $n-k$ of type 0, for each $k=0,\ldots,n$. 

Let us illustrate these bounds with some examples: 
\begin{itemize}
\item If $S=n!$ (i.e., if every permutation is a symmetry), then $\gcd(S,k!(n-k)!)=k!(n-k)!$ and both bounds become $n+1$. This agrees with our earlier observation that well-mixed haploid population models have $n+1$ reduced states.

\item If $S=1$, both bounds become $2^n$. This simply means that if there are no nontrivial symmetries among the $n$ sites, then there is no reduction and all $2^n$ states are non-equivalent.

\item A directed cycle graph (Fig.~\ref{fig:examples}f) has $S=n$. If $n$ is prime, then $\gcd(n,k!(n-k)!)=1$ for each $k=1,\ldots,n-1$.  So for prime $n$, both bounds coincide, yielding
\begin{equation}
    R=2+\frac{2^n-2}{n},
\end{equation}
in agreement with Eq.~\eqref{eq:reducedcycle}. The number of states is reduced by a factor of almost $n$, corresponding to the $n$ rotations of the cycle. 

\item For a star graph (Fig.~\ref{fig:examples}g), $S=(n-1)!$.  The lower bound is $n+1$ as long as $n\ge 3$. If $n$ is prime, then $\binom{n}{k}$ is divisible by $n$, meaning that $k!(n-k)!$ divides $(n-1)!$ and hence  $\gcd(S,k!(n-k)!)=k!(n-k)!$, for each $k=1, \ldots, n-1$.  Using this result, the upper bound evaluates to $2 + n(n-1)$ for prime $n$.  In this case, the bounds of $n+1 \le R \le 2+n(n-1)$ do not determine the value of $R$. However, considering the 2 possible states of the hub node, and the $n$ possible numbers of 1 alleles among the leaves, the number of reduced states is $R=2n$, which satisfies the lower and upper bounds.
\end{itemize}

\section{Discussion}

Symmetry arguments have been used since the earliest models of natural selection, first implicitly and in recent decades more explicitly \cite{Taylor,grafen2008natural,taylor2011groups,debarre2014social,mcavoy2015structural,mcavoy2015stochastic}. This work aims to formalize these arguments for a broad class of models, and provide a common language of definitions and foundational results. In particular, this work provides mathematical theory for how symmetry can reduce the states of the Markov chain representing selection. 

For most well-studied models of selection, such as those depicted in Figure \ref{fig:examples}, identifying the symmetries is straightforward.  However, for an arbitrary selection process, symmetries may be much more difficult to identify. Indeed, this problem includes, as a special case, the Graph Automorphism Problem---to determine whether a given graph has a nontrivial automorphism---which is NP-hard \cite{lubiw1981some}.

The framework employed in this work describes selection between two alleles, at a single genetic locus, in a population of fixed size and structure. This framework can be extended in a number of ways.  Generalizing to more than two alleles is relatively straightforward \cite{Allen2022Coalescent}: one can consider a set $A$ of allele types; the allele $x_g$ occupying site $g$ is then an element of $A$, and the state $\vx=(x_g)_{g \in G}$ becomes an element of $A^G$.  One must also specify a Markov chain on $A$ to characterize mutation of alleles. The definition of symmetry in Eq.~\eqref{eq:symmetry}, and the proofs of Lemma \ref{lem:group} and Theorem \ref{thm:sym_Markov}, carry over into the multi-allele setting. However, Theorem \ref{thm:numreduced} requires modification depending on number $|A|$ of alleles.

Generalizing to multiple genetic loci is also  straightforward---the set $G$ of sites can be expanded to include one site per locus for each haploid individual, or two sites per locus for each diploid. In most applications of multi-locus models, the loci play distinct roles, and therefore a symmetry must preserve the sites that correspond to each locus.

It is less obvious how to extend the definition of symmetry to dynamic population structures \cite{PachecoCoevolution,CorinaSet,PachecoCoevolution,wu2010evolution,Su2023StrategyEO}.  In this case, the state of the selection Markov chain consists not only of the population state $\vx$, but also the state of the population \emph{structure}, which can be represented by a variable $s$ belonging to some set of possibilities $\mathcal{S}$. The selection Markov chain must then include a transition rule for structures as well as population states. How can  symmetry be defined in this context?  One idea---used in a different context by McAvoy \cite{mcavoy2015stochastic}---is to require that the set $\mathcal{S}$ of structures  admit an action by the symmetric group $\Sym(G)$. This action characterizes how population structures change when sites are permuted.  Symmetry can then be represented by the subgroup of $\Sym(G)$ whose elements satisfy Eq.~\eqref{eq:symmetry} and whose action on $\mathcal{S}$ is compatible with the transition rule for population structures.


Symmetry, in the sense used here, is a property of \emph{models} of natural selection, rather than the manifestation of natural selection in the real world. 
Real-world biological processes rarely possess exact symmetry. Still, exploiting symmetry simplifies the analysis of  models, enabling tractable predictions for the real-world processes they represent. In this way, a formal theory of symmetry is not only of mathematical interest, but has practical value in aiding our understanding of natural selection.

\paragraph{Acknowledgements} I am grateful to Alex McAvoy for helpful conversations.
\paragraph{Competing interests} I declare no competing interests
\paragraph{Funding} This project was supported by Grant \#62220 from the John Templeton Foundation. Opinions expressed by the author do not necessarily reflect the views of the funding agency.

\appendix

\section{Mathematical proofs}
\label{app}

\begin{lemma}
\label{lem:group}
The symmetries, $\Sym(G,p)$, of a selection process, $(G,p)$, comprise a subgroup of the symmetric group $\Sym(G)$.
\end{lemma}

\begin{proof}
Since the elements of $\Sym(G,p)$ are permutations of $G$, $\Sym(G,p)$ is a subset of $\Sym(G)$.  To show that $\Sym(G,p)$ is a subgroup, it suffices to verify the following three properties: (i) the identity map on $G$ is a symmetry of $(G,p)$, (ii) if $\sigma$ and $\tau$ are symmetries, then $\sigma\circ \tau$ is as well; and (iii) if $\sigma$ is a symmetry then $\sigma^{-1}$ is as well. Property (i) follows directly from Eq.~\eqref{eq:symmetry}. Property (ii) is verfied as follows:
\begin{align*}
& \quad p_{\vx_{\sigma \circ \tau}} \big((\sigma \circ \tau)^{-1} \circ \alpha \circ (\sigma \circ \tau), (\sigma \circ \tau)^{-1} (U) \big)\\ 
& = 
p_{(\vx_\sigma)_\tau} \big(\tau^{-1} \circ (\sigma^{-1} \circ \alpha \circ \sigma) \circ \tau, \tau^{-1} \left( \sigma^{-1} (U) \right)\big)\\
& = 
p_{\vx_\sigma} \big(\sigma^{-1} \circ \alpha \circ \sigma, \sigma^{-1} (U) \big) & \text{by Eq.~\eqref{eq:symmetry}}\\
& = p_\vx(\alpha,U) & \text{by Eq.~\eqref{eq:symmetry} again.}
\end{align*}
Finally, property (iii) is verified as follows:
\begin{align*}
& \quad p_{\vx_{\sigma^{-1}}} \big(\left(\sigma^{-1}\right)^{-1} \circ \alpha \circ \sigma^{-1}, \left(\sigma^{-1}\right)^{-1}  (U) \big)\\ 
& = p_{\vx_{\sigma^{-1}}} \big(\sigma \circ \alpha \circ \sigma^{-1}, \sigma (U) \big)\\ 
& = 
p_{(\vx_{\sigma^{-1}})_\sigma} \big(\sigma^{-1} \circ (\sigma \circ \alpha \circ \sigma^{-1}) \circ \sigma, \sigma^{-1} \left( \sigma (U) \right)\big) & \text{by Eq.~\eqref{eq:symmetry}}\\
& = p_{\vx_{\sigma^{-1}\circ\sigma}} \big((\sigma^{-1} \circ \sigma) \circ \alpha \circ (\sigma^{-1} \circ \sigma), \sigma^{-1} \circ \sigma (U) \big)\\
& = p_\vx(\alpha,U). \qedhere
\end{align*}
\end{proof}

\setcounter{theorem}{0} 

\begin{theorem}
For any symmetry $\sigma \in \Sym(G,p)$ of a selection process $(G,p)$, and any states $\vx,\vy \in \{0,1\}^G$, $P_{\vx \to \vy} = P_{\vx_\sigma \to \vy_\sigma}$. 
\end{theorem}

\begin{proof}
Suppose that, in an arbitrary state $\vx$, parentage map $\alpha$ and mutation set $U$ occur, resulting in state $\vy$. By the transition rule for the selection Markov chain, we have, for each $g \in G$,
\begin{equation}
\label{eq:xytransition}
    y_g = \begin{cases} x_{\alpha(g)} & \text{if $g \notin U$}\\
    1- x_{\alpha(g)} & \text{if $g \in U$.}
    \end{cases}
\end{equation}
If $\sigma$ is a symmetry (or indeed any permutation of $G$), then substituting $\sigma(g)$ for $g$ yields
\begin{align*}
    y_{\sigma(g)}    & = \begin{cases} x_{\alpha\circ \sigma (g)} & \text{if $\sigma(g) \notin U$}\\
    1- x_{\alpha\circ \sigma (g)} & \text{if $\sigma(g) \in U$}
    \end{cases}\\
    & = \begin{cases} x_{\sigma \circ \sigma^{-1} \circ \alpha\circ \sigma (g)} & \text{if $g \notin \sigma^{-1}(U)$}\\
    1- x_{\sigma \circ \sigma^{-1} \circ \alpha\circ \sigma (g)} & \text{if $g \in \sigma^{-1}(U)$.}
    \end{cases}
\end{align*}
This means that $\vx_\sigma$ and $\vy_\sigma$ are related by
\begin{align}    
(\vy_\sigma)_g & = \begin{cases} (\vx_\sigma)_{\sigma^{-1} \circ \alpha\circ \sigma (g)} & \text{if $g \notin \sigma^{-1}(U)$}\\
    1- (\vx_\sigma)_{\sigma^{-1} \circ \alpha\circ \sigma (g)} & \text{if $g \in \sigma^{-1}(U)$.}
    \end{cases}
\end{align}
It follows that if parentage map $\sigma^{-1} \circ \alpha \circ \sigma$ and mutation set $\sigma^{-1}(U)$ occur in state $\vx_\sigma$, the resulting state is $\vy_\sigma$.  The symmetry property, Eq.~\eqref{eq:symmetry}, then implies that $P_{\vx \to \vy} = P_{\vx_\sigma \to \vy_\sigma}$.
\end{proof}

\begin{theorem}
\label{thm:numreduced}
For a selection process $(G,p)$, let $n=|G|$ be the number of states, $S=|\Sym(G,p)|$ be the number of symmetries, and $R$ be the number of states in the reduced chain $\mathcal{R}(G,p)$. Then $R$ is bounded by 
\begin{equation}
\label{eq:numreduced}
    \max\left(n+1, 2 + \frac{2^{n}-2}{S} \right) \le R \le 2 + \frac{1}{S}\sum_{k=1}^{n-1} \binom{n}{k} \gcd(S,k!(n-k)!).
\end{equation}
\end{theorem}

\begin{proof}
By the definition of $\mathcal{R}(G,p)$, $R$ is the number of orbits for the action of $\Sym(G,p)$ on $\{0,1\}^G$. Applying the orbit-counting theorem, also known as Burnside's Lemma (e.g., Ref.~\cite{armstrong1997groups}, Chapter 17), yields
\begin{equation}
\label{eq:Burnside}
    R = \frac{1}{S} \sum_{\vx \in \{0,1\}^G} |\Sym(G,p;\vx)|.
\end{equation}
Above, $\Sym(G,p;\vx)$ is the stabilizer of $\vx$, that is, the subgroup of $\Sym(G,p)$ that leaves $\vx$ fixed:
\begin{equation}
    \Sym(G,p;\vx) = \{\sigma \in \Sym(G,p) \mid \vx_\sigma = \vx\}.
\end{equation}
The sum in Eq.~\eqref{eq:Burnside} can be decomposed according to the number of 1 alleles:
\begin{equation}
\label{eq:Rsum1}
    R = \frac{1}{S}\sum_{k=0}^{n}\; \sum_{\vx: \Sigma \vx = k} |\Sym(G,p;\vx)|.
\end{equation}
Since any symmetry leaves the single-allele states $\mathbf{0}$ and $\mathbf{1}$ fixed, we have $|\Sym(G,p;\vzero)|=|\Sym(G,p;\vone)|=S$.  The $k=0$ and $k=n$ terms of the outer sum therefore evaluate to $S$, and Eq.~\eqref{eq:Rsum1} becomes
\begin{equation}
\label{eq:Rsum2}
    R = 2+\frac{1}{S}\sum_{k=1}^{n-1}\; \sum_{\vx: \Sigma \vx = k} |\Sym(G,p;\vx)|.
\end{equation}
For a given state $\vx$, any element of $\Sym(G,p;\vx)$ must preserve the set $\{g \in G \mid x_g=0\}$ of sites containing allele 0, as well as the set $\{g \in G \mid x_g=1\}$ of sites containing 1. This means that $\Sym(G,p;\vx)$ is a subgroup of $\Sym(\{g \in G \mid x_g=0\})\times \Sym(\{g \in G \mid x_g=1\})$.  Since $\Sym(G,p;\vx)$ is also a subgroup of $\Sym(G,p)$, the order of $\Sym(G,p;\vx)$ must divide that of both $\Sym(G,p)$ and $\Sym(\{g \in G \mid x_g=0\})\times \Sym(\{g \in G \mid x_g=1\})$ by Lagrange's Theorem (e.g., \cite{armstrong1997groups}, Chapter 11).  It follows that $|\Sym(G,p;\vx)|$ is a common divisor of $S$ and $k!(n-k)!$. In particular, we have the bounds
\begin{equation}
    1 \le |\Sym(G,p;\vx)| \le \gcd(S,k!(n-k)!).
\end{equation}
Applying these bounds to Eq.~\eqref{eq:Rsum2}, and noting that there are $\binom{n}{k}$ states $\vx$ with $\Sigma \vx = k$, we obtain 
\begin{equation}
2 + \frac{2^{n}-2}{S} \le R \le 2 + \frac{1}{S}\sum_{k=1}^{n-1} \binom{n}{k} \gcd(S,k!(n-k)!).
\end{equation}
Finally, since states in the same orbit have the same number of each allele, there must be at least $n+1$ orbits (for each value of $\Sigma \vx \in \{0, \ldots, n\}$), providing the additional lower bound in Eq.~\eqref{eq:numreduced}.
\end{proof}

\bibliographystyle{vancouver}
\bibliography{arbitrary}

\end{document}